\documentclass[aip,pof,preprint]{revtex4-1}
\usepackage{epsfig,graphics,amssymb,amsmath,subeqnarray}
\usepackage{graphicx}
\usepackage{caption}

\usepackage{color}
\usepackage{array, booktabs}
\usepackage{float}
\restylefloat{table}
\usepackage{multirow}

\def\u{{\mathbf{u}}}

\def\v{{\mathbf{v}}}

\def\eh{{\mathbf{e}}}

\def\sigmaB{{\boldsymbol{\sigma}}}

\def\gammaB{{\boldsymbol{\gamma}}}

\def\I{{\textbf{I}}}

\def\tauB{{\boldsymbol{\tau}}}

\def\varepsilon{\epsilon}

\def\DT{{\text{D}}}
\def\ST{{\text{S}}}

\def\tr{{\text{tr}}}

\begin{document}
\title{A note on the breathing mode of an elastic sphere in Newtonian and complex fluids}

\author{Vahe Galstyan}
\affiliation{Department of Physics, Columbia University, New York, NY 10027}

\author{On Shun Pak}
\affiliation{
Department of Mechanical Engineering, 
Santa Clara University, Santa Clara, CA 95053}
\affiliation{
Department of Mechanical and Aerospace Engineering, 
Princeton University, Princeton, NJ 08544}

\author{Howard A. Stone}
\affiliation{
Department of Mechanical and Aerospace Engineering, 
Princeton University, Princeton, NJ 08544}

%\date{\today}

\begin{abstract}
Experiments on the acoustic vibrations of elastic nanostructures in fluid media have been used to study the mechanical properties of materials, as well as for mechanical and biological sensing. The medium surrounding the nanostructure is typically modeled as a Newtonian fluid. A recent experiment however suggested that high-frequency longitudinal vibration of bipyramidal nanoparticles could trigger a viscoelastic response in water-glycerol mixtures [M. Pelton \textit{et al.}, ``Viscoelastic flows in simple liquids generated by vibrating nanostructures,'' Phys. Rev. Lett. \textbf{111}, 244502 (2013)]. Motivated by these experimental studies, we first revisit a classical continuum mechanics problem of the purely radial vibration of an elastic sphere, also called the breathing mode, in a compressible viscous fluid, and then extend our analysis to a viscoelastic medium using the Maxwell fluid model. The effects of fluid compressibility and viscoelasticity are discussed. Although in the case of longitudinal vibration of  bipyramidal nanoparticles, the effects of fluid compressibility were shown to be negligible, we demonstrate that it plays a significant role in the breathing mode of an elastic sphere. On the other hand, despite the different vibration modes, the breathing mode of a sphere triggers a viscoelastic response in water-glycerol mixtures similar to that triggered by the longitudinal vibration of bipyramidal nanoparticles. We also comment on the effect of fluid viscoelasticity on the idea of destroying virus particles by acoustic resonance.

%\textcolor{blue}{We also investigate the effect of the fluid viscoelasticity on the lifetime of the breathing mode of nanometer scale viruses which are typically modeled as elastic spheres. We find that in highly viscous media, such as glycerol, viscoelasticity, as an energy storing mechanism, can substantially increase the vibration lifetime of viruses, which is an important criterion for destroying viruses though resonance.}
\end{abstract}

\maketitle

\section{Introduction}

Studies on the vibration of elastic nanoparticles embedded in fluid media have attracted considerable attention recently, due to potential applications, for example, as an alternative nondestructive tool for characterizing material properties\cite{portales} and designing mechanical and biological sensors.\cite{jensen, verbridge, arlett} Low damping, \textit{i.e.}, a high quality factor, is desirable in these applications for high detection sensitivity. It is therefore of interest to investigate the damping mechanisms due to energy dissipation or energy transfer to the surrounding media. In addition, biological nanoparticles such as viruses have also been modeled as elastic spheres in the studies of their vibration characteristics in different media, motivated by the idea of destroying viruses in a living host with ultrasound waves via resonance. \cite{babincova, ford,sav_murray, talati, stephanidis}

%\textcolor{red}{In addition, the idea of destroying viruses in a living host with ultrasound waves via resonance has motivated interest in the vibration characteristics of viruses, which have been modeled as elastic spheres. \cite{ford,sav_murray, talati, stephanidis} }

%\textcolor{red}{In addition, biological nanoparticles such as viruses have been modeled as elastic spheres \cite{ford,sav_murray, talati}, and their vibration characteristics  

The resonant frequencies and damping characteristics of mechanical nanostructures with various shapes have been measured by a variety of experimental techniques.\cite{hartland, nano, fujii} From the modeling perspective, acoustic vibrations of elastic bodies are classical problems in continuum mechanics. For example, Lamb  \cite{lamb} studied theoretically the vibrations of an elastic sphere in vacuum. Subsequent works considered the vibration modes of elastic structures with different geometries, which are summarized briefly in Table \ref{table:authors}. In addition, the effect of different surrounding environments, including an elastic solid matrix,\cite{dub} and inviscid \cite{kheisin} and viscous \cite{sav, chakraborty} fluid media, were also considered in later studies (Table \ref{table:authors}). As a remark, for a nanoparticle with a typical size of tens of nanometers, the atomic spacing is usually sufficiently small that a continuum description is valid. \cite{sav,chakraborty} Such a continuum approach was also shown to be successful in predicting the resonant frequency (on the order of tens of GHz) of a gold nanoparticle vibrating in water. \cite{chakraborty}

\begin{table} [h]
\scriptsize
\begin{tabular*}{\textwidth} { >{\centering\arraybackslash} m{2.7cm}  | >{\centering\arraybackslash} m{2.5cm} | >{\centering\arraybackslash} m{1.5cm} | >{\centering\arraybackslash} m{2.8cm} | >{\centering\arraybackslash} m{2.7cm} | >{\centering\arraybackslash} m{3.6cm} }
\hline

 & \multirow{2}{*} {Elastic Structure} & \multicolumn{3}{c|}{Surrounding Medium} & \multirow{2}{*} {Vibration Modes} \\ \cline{3-5}
 & & Viscosity & Compressibility & Rheology & \\ \hline \hline

Lamb\cite{lamb}  (1882)& Sphere & \multicolumn{3}{c|}{Vacuum} &  Torsional \& spheroidal\\ \hline
Kheisin\cite{kheisin} (1967) & Sphere & Inviscid & Compressible & Newtonian fluid & Breathing \\ \hline
Dubrovskiy \& Morochnik \cite{dub} (1981)& Sphere & \multicolumn{3}{c|}{Elastic solid matrix} &  Torsional \& spheroidal\\ \hline

Saviot \textit{et al.} \cite{sav} (2007)& Sphere & Viscous & Compressible & Newtonian fluid &  Torsional \& spheroidal\\ \hline

Chakraborty \textit{et al.} \cite{chakraborty} (2013)& Circular cylinder Conical cylinder Bipyramid & Viscous & Incompressible & Newtonian fluid & Longitudinal vibrations along the major axis\\ \hline

Pelton \textit{et al.} \cite{pelton} (2013)& Bipyramid & Viscous & Compressible \& incompressible & Viscoelastic fluid (Maxwell model) & Longitudinal vibrations along the major axis\\ \hline

This work & Sphere & Viscous & Compressible \& incompressible & Viscoelastic fluid (Maxwell model) & Breathing \\ \hline
\end{tabular*}
\caption{Some theoretical studies of the vibrations of elastic structures in different media.}
\label{table:authors}
\end{table}

Recently, an experiment on the vibration of a bipyramidal gold nanoparticle (in the shape of a pair of truncated cones) in water-glycerol mixtures suggested that the high-frequency (20 GHz) vibration could trigger viscoelastic responses in the mixture, even in small molecule liquids.\cite{pelton} The bipyramidal nanoparticle was excited to vibrate longitudinally along its major axis in a water-glycerol mixture. When increasing the glycerol mass fraction, the quality factor of the vibration displayed non-monotonic variations not explained by a Newtonian fluid model; a viscoelastic fluid model (linear Maxwell model) however captured the behavior. In contrast to the longitudinal vibrations of bipyramidal nanoparticles, the breathing mode, which refers to purely radial vibrations, is mainly excited for spherical nanoparticles,\cite{nano} whose behavior in a viscoelastic fluid medium remains unexplored. In this paper, we first revisit the classical problem of the breathing mode of a vibrating elastic sphere in a Newtonian fluid, and then extend our studies to complex fluid media. 

We organize the paper by first presenting a continuum mechanics formulation in Sec.~\ref{sec:formulation} for the elasticity problem of a radially vibrating sphere (Sec.~\ref{sec:elasticity}) and the propagation of acoustic waves in the fluid medium (Sec.~\ref{sec:fluid}). The two problems are then coupled by matching the velocities and stresses at the interface of the solid (Sec.~\ref{sec:coupling}). An analytical eigenvalue equation determining the vibration frequencies of an elastic sphere in a compressible, viscous, Newtonian fluid medium is obtained in Sec.~\ref{sec:results}. The results are validated against previous theoretical and experimental studies of a vibrating gold nanosphere in water (Sec.~\ref{sec:goldwater}), followed by parametric studies of the quality factor (Sec.~\ref{sec:parametric}). The calculations are then extended to the viscoelastic case in Sec.~\ref{sec:viscoelastic}, with results and remarks discussed in Sec.~\ref{sec:discussion}.

\section{Formulation} \label{sec:formulation}

\begin{figure}[h]
\centering
\includegraphics[width=0.35\textwidth]{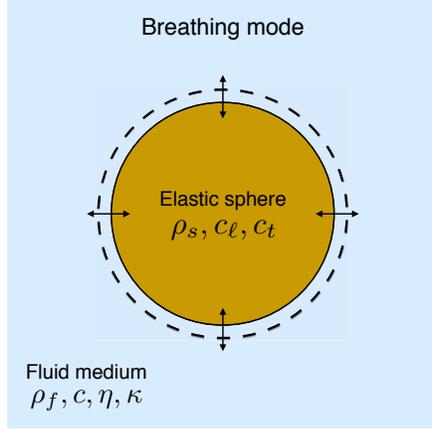}
  \caption{Schematic representation of the breathing mode (purely radial vibration) of an elastic sphere (density $\rho_s$, longitudinal wave speed $c_\ell$, and transverse wave speed $c_t$) in a fluid medium (density $\rho_f$, wave speed $c$, shear viscosity $\eta$, and bulk viscosity $\kappa$).}
        \label{fig:formulation}
\end{figure}

We consider the purely radial vibration of an elastic sphere of radius $R$ in a compressible viscous fluid. This spherically symmetric motion is also called the breathing mode (see Fig. \ref{fig:formulation} for a schematic representation of the problem). The displacement field of the elastic sphere, $\u$, is governed by the Navier equation in elasticity
\begin{align}
\rho_s \frac{\partial^2 \u}{\partial t^2} = \mu \nabla^2 \u +(\mu+\lambda) \nabla (\nabla \cdot \u), \label{eqn:navier}
\end{align}
where $\rho_s$ is the density of the solid, and $\mu$ and $\lambda$ are the Lam\'{e} elastic parameters. We consider small-amplitude acoustic waves in the fluid, and hence the velocity field, $\v$, is governed by the linearized Navier-Stokes equation for compressible flows
\begin{align}
\rho_f \frac{\partial \v}{\partial t} = - \nabla p + \eta \nabla^2 \v + \left(\kappa + \frac{\eta}{3}\right) \nabla (\nabla \cdot \v) \label{eqn:navier-stokes},
\end{align}
where $\rho_f$ is the density of the fluid, $\eta$ is the shear viscosity, $\kappa$ is the bulk viscosity, and $p$ is the thermodynamic pressure.

Since the vibration is purely radial, we utilize a spherical coordinate system located at the center of the sphere. The displacement field of the elastic sphere, $\u(r,t) = u(r,t) \eh_r$, and the velocity field of the fluid, $\v(r,t)=v(r,t) \eh_r$, have only radial components that are functions of the distance from the origin $r$ and time $t$. With this geometrical symmetry, the identity  $\nabla^2 \mathbf{a} \equiv \nabla (\nabla \cdot \mathbf{a})$ holds for a vector field $\mathbf{a}$, which simplifies (\ref{eqn:navier}) and (\ref{eqn:navier-stokes}), respectively, to
\begin{subequations}
\begin{align}
\rho_s \frac{\partial^2 \u}{\partial t^2} &= (2\mu+\lambda) \nabla^2 \u, \label{eqn:navier2}\\
\rho_f \frac{\partial \v}{\partial t} &= - \nabla p  +\beta \nabla^2 \v \label{eqn:navier-stokes2},
\end{align}
\end{subequations}
where $\beta =\kappa + 4\eta/3$.

\subsection{Elasticity: radial vibration of an elastic sphere} \label{sec:elasticity}
We take the divergence of (\ref{eqn:navier2}) and define the scalar function $\phi = \nabla \cdot \u$ to obtain
\begin{align}
\frac{\partial^2  \phi}{\partial t^2} = c_\ell^2 \nabla^2 \phi, \label{eqn:Navier}
\end{align}
where $c_\ell = \sqrt{(2\mu+\lambda)/\rho_s}$, which physically represents the velocity of longitudinal waves in an elastic material. We find time-periodic solutions of the form $\phi(r,t) = \Phi(r) e^{-i \omega t}$ for the breathing mode. By separation of variables we obtain
\begin{align}
\phi(r,t) = \sum_{n=1}^\infty A_n \frac{\sin (k_{s,n} r )}{r} e^{-i \omega_n t},
\end{align}
where $A_n$ are arbitrary constants, and $k_{s,n}=\omega_n/c_\ell$ are the unknown eigenvalues. The displacement field of the elastic sphere can then be determined by direct integration using the definition $\phi=\nabla \cdot \u =(1/r^2)\partial(r^2 u)/\partial r$, which results in
\begin{align}
u(r,t) = \sum_{n=1}^\infty A_n \left[ \frac{\sin (k_{s,n} r)}{k_{s,n}^2 r^2} - \frac{\cos(k_{s,n} r)}{k_{s,n} r} \right] e^{-i \omega_n t} . \label{eqn:u}
\end{align}

\subsection{Fluid dynamics: propagation of acoustic waves}\label{sec:fluid}
We now turn to the  propagation of small-amplitude acoustic waves in the fluid surrounding the vibrating sphere. The linearized continuity equation for a compressible fluid is given by
\begin{align}
\frac{\partial \rho'}{ \partial t} + \rho_f \nabla \cdot \v = 0, \label{eqn:continuity}
\end{align}
where $\rho'(r,t)$ represents the density fluctuation; $\mid \rho' \mid \ll \rho_f$. Together with the equation of state, $p = c^2 \rho'$, where $c$ is the speed of sound in the fluid, (\ref{eqn:continuity}) can be combined with (\ref{eqn:navier-stokes2}) to arrive at an equation for the density fluctuations
\begin{align}
\frac{\partial^2 \rho'}{\partial t^2} = c^2 \nabla^2 \rho' + \frac{\beta}{\rho_f} \nabla^2 \frac{\partial \rho'}{\partial t} \cdot \label{eqn:rhoEq}
\end{align}
We again seek time-periodic solutions with the same vibration frequencies of the elastic sphere. Then, (\ref{eqn:rhoEq}) can be solved by separation of variables to yield
\begin{align}
\rho'(r,t) = \rho_f \sum_{n=1}^\infty \tilde{A}_n  \frac{e^{i(k_{f,n}  r - \omega_n t)}}{r},
\end{align}
where $\tilde{A}_n$ are arbitrary constants and $k_{f,n}= \frac{\omega_n}{c \sqrt{1-i \omega_n \beta/\rho_fc^2}}$. The corresponding velocity field can then be obtained from the continuity equation (\ref{eqn:continuity}) as
\begin{align}
 v(r,t) = \sum_{n=1}^\infty \tilde{A}_n i \omega_n  \frac{(1-i k_{f,n} r)}{k_{f,n}^2 r^2} e^{i(k_{f,n} r - \omega_n t)}  . \label{eqn:FluidVelocity}
\end{align}

%The pressure field is given by
%\begin{align}
%p = c^2 \rho' = \sum_{n=1}^\infty \tilde{b}_n c^2 \rho_f \frac{e^{i(k_{f,n}  r - \omega_n t)}}{r} \cdot
%\end{align}

\subsection{Elastohydrodynamics: coupling the radial vibration of an elastic sphere to the surrounding fluid} \label{sec:coupling}

We couple the vibration of the elastic sphere to the fluid by matching the velocities and normal stresses at the boundary between the fluid and the vibrating sphere. Since small amplitude vibrations are considered, we use domain perturbation and expand the velocities and stresses at the boundary about the equilibrium radius of the sphere, $R$, a constant, keeping only the leading-order terms. For continuity of velocity, we compute the time derivative of the displacement field of the elastic sphere (\ref{eqn:u}) and equate it to the velocity field in the fluid (\ref{eqn:FluidVelocity}) evaluated at $r=R$, which results in
\begin{align}
 - A_n  \left[ \frac{\sin (k_{s,n} R)}{k_{s,n}^2 R^2} - \frac{\cos(k_{s,n} R)}{k_{s,n} R} \right] =  \tilde{A}_n  \frac{(1-i k_{f,n} R)}{k_{f,n}^2 R^2} e^{i k_{f,n} R}  . \label{eqn:VelMatch}
\end{align}

The stress tensor in the solid is given by 
\begin{align}
\sigmaB^s = \lambda (\nabla \cdot \u) \I + 2\mu \gammaB,
\end{align}
where $\gammaB = \frac{1}{2}(\nabla \u + (\nabla\u)^T)$ is the strain tensor. For the spherically symmetric case considered here, the only non-zero component of the stress tensor is 
\begin{align}
\sigma^s_{rr} (r,t)=  \sum_{n=1}^\infty A_n  \left\{ \frac{\lambda+2\mu}{r} \sin(k_{s,n} r) - \frac{4\mu}{r} \left( \frac{ \sin(k_{s,n} r)}{k_{s,n}^2 r^2} - \frac{\cos(k_{s,n} r) }{k_{s,n} r} \right) \right\} e^{-i\omega_n t} . \label{eqn:SolidStress}
\end{align}
The stress tensor in the fluid is given by 
\begin{align}
\sigmaB^f = \left(-p + \kappa~ \tr(\dot{\gammaB}) \right) \I + 2\eta \left( \dot{\gammaB}- \frac{\tr(\dot{\gammaB})}{3}\I \right), \label{eqn:Newtonian}
\end{align}
 where $\tr(\dot{\gammaB}) = \nabla \cdot \v$, and $\dot{\gammaB}=\frac{1}{2}(\nabla \v + (\nabla\v)^T)$ represents the rate-of-strain tensor. The only non-zero component of the stress tensor is 
\begin{align}
\sigma^f_{rr} (r,t) = \sum_{n=1}^\infty \tilde{A}_n \frac{e^{i (k_{f,n} r-\omega_n t)}}{r} \left( - \rho_f c^2 + i \beta \omega_n - 4\eta i \omega_n\frac{( 1 - i k_{f,n} r)}{k_{f,n}^2 r^2 }   \right), \label{eqn:FluidStress}
\end{align}
where the equation of state and the continuity equation have been used. 

Evaluating and matching the stresses in the solid (\ref{eqn:SolidStress}) and fluid (\ref{eqn:FluidStress}) at $r=R$ leads to 
\begin{gather}
A_n  \left( (\lambda+2\mu) \sin(k_{s,n} R) - 4\mu \left[ \frac{ \sin(k_{s,n} R)}{k_{s,n}^2 R^2} - \frac{\cos(k_{s,n} R) }{k_{s,n} R} \right] \right)  \notag \\
= \tilde{A}_n \left( - \rho_f c^2 + i \beta \omega_n - 4\eta i \omega_n\frac{( 1 - i k_{f,n} R)}{k_{f,n}^2 R^2} \right) e^{i k_{f,n} R} . \label{eqn:StressMatch}
\end{gather}

\section{Results} \label{sec:results}

For non-trivial solutions to the system of equations formed by equations (\ref{eqn:VelMatch}) and (\ref{eqn:StressMatch}), we require the determinant of the matrix representing this system to vanish, which leads to the eigenvalue equation for the natural frequencies $\omega_n$
\begin{align}
 k_{s,n}^2 R^2  {\bigg[\frac{\rho_f}{\rho_s (1 - i k_{f,n}R)} + \bigg( \frac{k_{s,n} R}{\tan(k_{s,n}R)}-1 \bigg)^{-1}\bigg]} +\frac{4i\eta k_{s,n}}{\rho_s c_\ell} + 4 \left(\frac{c_t}{c_\ell}\right)^2 = 0. \label{eqn:eigen_newton}
\end{align}
We note that (\ref{eqn:eigen_newton}) is a transcendental equation to be solved numerically. Here $c_t = \sqrt{\mu/\rho_s}$ represents the velocity of transverse elastic waves. We calculate only the results for the fundamental mode ($n=1$) here, since it is mainly the mode detected in experimental measurements; \cite{nano} higher order modes can be obtained by finding other roots of the same equation. Hereafter, for simplicity we denote $\omega =\omega_1$, $k_s=k_{s,1} =\omega/c_\ell$, and $k_{f}=k_{f,1} = \frac{\omega}{c\sqrt{1-i\omega \beta/\rho_f c^2}}$.

For an inviscid flow ($\kappa=\eta=0$), the eigenvalue condition (\ref{eqn:eigen_newton}) for $\omega$ reduces to
\begin{align}
\frac{\rho_f}{\rho_s} = \left[ \frac{4}{k_s^2 R^2} \left(\frac{c_t}{c_\ell}\right)^2 +  \left( \frac{k_s R}{\tan(k_s R)} -1\right)^{-1}   \right] (i\omega R/c-1), \label{eqn:inviscid}
\end{align}
which was previously obtained by Kheisin. \cite{kheisin}  The case $\rho_f =0$ corresponds to an elastic sphere vibrating in a vacuum, \cite{lamb} where the eigenvalue condition further simplifies to
\begin{align}
\frac{k_s R}{\tan(k_s R)} =  1- \frac{k_s^2 R^2}{4} \left(\frac{c_\ell}{c_t} \right)^2. \label{eqn:vacuum}
\end{align}
Dubrovsky and Morochnik \cite{dub} considered the breathing mode of an elastic sphere surrounded by another elastic medium. The eigenvalue conditions (\ref{eqn:eigen_newton})--(\ref{eqn:vacuum}) can be alternatively obtained from the inviscid results of Dubrovsky and Morochnik by modifying the constitutive relation to include the viscous contribution from the fluid.\cite{sav}

For the case of vacuum surrounding the vibrating sphere, the frequencies determined from (\ref{eqn:vacuum}) are real because there is no damping outside the sphere. When the vacuum is replaced by an inviscid fluid medium, the frequencies determined from the roots of (\ref{eqn:inviscid}) become complex, which implies that the free oscillation of the elastic sphere in a fluid takes the form of a sinusoid attenuating exponentially. \cite{kheisin}

\subsection{Breathing mode of a gold nanosphere in a Newtonian fluid} \label{sec:goldwater}

Here we first consider the experimental setup of Ruijgrok \textit{et al.},\cite{nano} where the breathing mode of a single gold nanosphere of 40 nm radius in water was measured. We denote the complex frequency $\omega = \omega_r +i \omega_i$, where $\omega_r$ and $\omega_i$ are, respectively, the real and imaginary parts of the frequency. In this paper we define the quality factor as ${Q = -\sqrt{\omega_r^2+\omega_i^2} /(2 \omega_i)}$, which is the same as the definition adopted in Pelton \textit{et al.} \cite{pelton} Note that Ruijgrok \textit{et al.}\cite{nano} defined the quality factor differently as $Q'=-\omega_r/(2\omega_i)$, which gives only small quantitative differences for the vibrations with low damping ($\omega_i/\omega_r \ll 1$) considered in this section. With the material constants of water given in Table \ref{table:parameters}, \footnote{The values of material parameters used in Ruijgrok \textit{et al.}\cite{nano} are slightly different from those in Table \ref{table:parameters}, which result in the slight differences in the calculated quality factors} in the inviscid limit a quality factor $Q= 53.7$ is obtained using (\ref{eqn:inviscid}). \cite{nano} Note that the attenuation in oscillation is not due to viscous dissipation in the fluid since an inviscid medium is considered. Instead, the attenuation comes from the propagation of energy into the surrounding medium away from the vibrating sphere. Compressibility in the fluid provides a mechanism for energy to propagate away from the vibrating sphere through acoustic waves. When the shear and bulk viscosities of water are taken into account, (\ref{eqn:eigen_newton}) gives a slightly reduced quality factor of $Q=52.6$, as reported by Ruijgrok \textit{et al.}\cite{nano} 

Although the modification of the quality factor by the viscous effects is relatively minor in this case, a more significant reduction in the quality factor can occur when water is replaced by a water-glycerol mixture, which is the fluid medium considered by Pelton \textit{et al.} \cite{pelton} The quality factor drops from $Q=40.6$ (inviscid) to $Q=33.4$ (viscous) for the vibration of the same gold nanosphere in a water-glycerol mixture with the glycerol mass fraction $\chi =0.56$, where the mixture is modeled as a Newtonian fluid. The parameter values and quality factors of these theoretical predictions are summarized in Table \ref{table:parameters}. The viscous effects are discussed in terms of dimensionless variables in Sec.~\ref{sec:parametric}.

\begin{table} [h]
\scriptsize

\begin{tabular*}{\textwidth} { >{\centering\arraybackslash} p{3.0cm}  | >{\centering\arraybackslash} p{5cm} | >{\centering\arraybackslash} p{2.5cm} | >{\centering\arraybackslash} p{2.5cm} | >{\centering\arraybackslash} p{2.7cm} }
\hline
\multirow{2}{*} {Gold nanoparticle} & \multirow{2}{*} {Surrounding medium}  & \multicolumn{3}{c}{Predicted quality factor, $Q$} \\ \cline{3-5}

& & Inviscid & Viscous & Viscoelastic \\ \hline \hline

&\underline{Water:} &  &  & \\ 

& $\rho_f =  1000$ kg/m$^{3}$, $c=1510$ m/s & 53.7 & 52.6 & 52.2 \\

 $\rho_s = 19700$ kg/m$^{3}$ & $\eta = 0.000894$ Pa$\cdot$s, $\kappa=0.00286$ Pa$\cdot$s& ($f=37.9$ GHz) & ($f=37.8$ GHz) & ($f=37.8$ GHz) \\ 
 
 $c_\ell = 3240$ m/s  & $\lambda=0.647$ ps (for viscoelastic case)  & & & \\ \cline{2-5}
 
 $c_t=1200$ m/s & \underline{Water-glycerol mixture ($\chi=0.56$):}  & & & \\
 
& $\rho_f =  1140$ kg/m$^{3}$, $c=1760$ m/s& 40.6 & 33.4 & 34.5 \\
 
& $\eta = 0.00527$ Pa$\cdot$s, $\kappa=0.0116$ Pa$\cdot$s& ($f=37.9$ GHz) & ($f=37.6$ GHz) & ($f=37.7$ GHz) \\ 

& $\lambda=3.51$ ps (for viscoelastic case) & & & \\ \hline

\end{tabular*}
\caption{Theoretical predictions of the quality factor $Q$ and vibration frequency $f = \omega_r/(2\pi) $ of the breathing mode of a single gold nanosphere with $40$ nm radius vibrating in inviscid, viscous, and viscoelastic fluid media. Refer to Appendix \ref{sec:data} for the details of the material constants.}
\label{table:parameters}
\end{table}

The natural frequency of an elastic sphere oscillating in an incompressible flow can be obtained by using the corresponding continuity equation, $\nabla \cdot \v = 0$ (see Appendix \ref{appendix:incompress}), or by simply taking the wave speed in the fluid to infinity ($c \rightarrow \infty$) in (\ref{eqn:eigen_newton}), which leads to

\begin{align}
 k_{s}^2 R^2  {\bigg[\frac{\rho_f}{\rho_s} + \bigg( \frac{k_{s} R}{\tan(k_{s}R)}-1 \bigg)^{-1}\bigg]} +\frac{4i\eta k_{s}}{\rho_s c_\ell} + 4 \left(\frac{c_t}{c_\ell}\right)^2 = 0. \label{eqn:eigen_newton_incompress}
\end{align}

%The derivation of (\ref{eqn:eigen_newton_incompress}) from the fluid incompressibility condition ($\nabla \cdot \v = 0$) is given in Appendix \ref{appendix:incompress}. 
Using the material constants of water from Table \ref{table:parameters}, we obtain a quality factor $Q=841$ for the incompressible viscous case, which is significantly larger than the quality factor when compressibility is taken into account ($Q=52.6$). Although Pelton \textit{et al.} \cite{pelton} showed numerically that the effect of compressibility is insignificant for the longitudinal oscillations of a bipyramidal gold nanoparticle (with 25 nm base diameter and 90 nm total height), we demonstrate here an example of the breathing mode of nanospheres where fluid compressibility should be considered to obtain a reasonable prediction of the quality factor.

\subsection{Parametric studies} \label{sec:parametric}

We study the effects of different physical parameters on the quality factor by nondimensionalizing (\ref{eqn:eigen_newton}) to identify the dimensionless groups in our problem: 
\begin{align}
\label{eqn:nondimen}
s = \frac{\omega R}{c_\ell}, \quad
\tilde{\rho} = \frac{\rho_f}{\rho_s},  \quad
\tilde{c}_t = \frac{c_t}{c_\ell},  \quad
\tilde{c} = \frac{c}{c_\ell},  \quad
\tilde{R} = \frac{\rho_f c R}{\eta},  \quad
\alpha  = \frac{\kappa}{\eta}.
\end{align}

The dimensionless form of the eigenvalue equation (\ref{eqn:eigen_newton}) is
\begin{align}
 s^2 \left[ \frac{\tilde{\rho}}{1 - i\tilde{k}_f} + \left( \frac{s}{\tan s} - 1 \right)^{-1}\right] + \frac{4i s \tilde{c} \tilde{\rho}}{\tilde{R}} + 4 \tilde{c}^2_t = 0, 
\end{align}
where
\begin{align}
\tilde{k}_f(s) = \frac{s/\tilde{c}}{\sqrt{1 - i\left(\alpha+\frac{4}{3} \right)s / (\tilde{R} \tilde{c} ) }} \cdot
\end{align}

We first examine the relative importance of the viscous dissipation mechanism compared with the damping caused by the propagation of acoustic waves away from the elastic nanosphere at different values of $\tilde{R}=\rho_f c R/\eta$. The relative importance is characterized by comparing the quality factor when viscous effects are considered $Q_\text{visc}$, to the quality factor when the fluid medium is assumed to be inviscid $Q_\text{inv}$ (hence, all damping is due to the propagation of acoustic waves away from the source). The ratio $Q_\text{visc}/Q_\text{inv}$ should approach unity when viscous effects are negligible and decrease as the viscous effects become significant.

\begin{figure}[h]
\centering
 \includegraphics[width=0.35\textwidth]{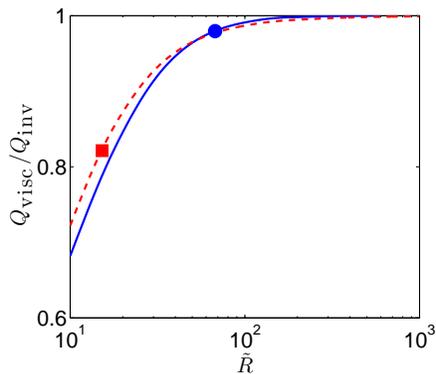}
 \caption{Parametric study of the ratio $Q_\text{visc}/Q_{\text{inv}}$ for different values of $\tilde{R}$ using the Newtonian fluid model. The symbols represent the vibrations of a gold nanosphere with $40$ nm radius in pure water (blue dot) and water-glycerol mixture (glycerol mass fraction $\chi=0.56$, red square). The blue solid (red dotted) line is obtained by varying $\tilde{R}$ with all other dimensionless parameters ($\tilde{\rho}$, $\tilde{c_t}$, $\tilde{c}$ and $\alpha$) kept the same as those represented by the blue dot (red square). The dimensionless parameters (see equation (\ref{eqn:nondimen})) are obtained from the material constants presented in Table \ref{table:parameters}.}
        \label{fig:re_dependence}
\end{figure}

For the case of a gold nanosphere with $40$ nm radius vibrating in water, we obtain the corresponding values of $\tilde{R} = 67.6$ and the ratio $Q_\text{visc}/Q_{\text{inv}}\approx0.98$ (blue dot, Fig.~{\ref{fig:re_dependence}}). Viscous dissipation is therefore negligible as reported by Ruijgrok \textit{et al.}\cite{nano} However, the viscous effects become more significant for the same vibration in a water-glycerol mixture with glycerol mass fraction $\chi=0.56$. (see Table \ref{table:parameters} for the properties of the mixture \cite{slie}). The corresponding value of $\tilde{R} = 15.2$, which results in $Q_\text{visc}/Q_{\text{inv}}\approx0.82$ (red square, Fig.~\ref{fig:re_dependence}). The blue solid line (red dotted line) is obtained by varying $\tilde{R}$ while keeping all other dimensionless parameters ($\tilde{\rho}$, $\tilde{c_t}$, $\tilde{c}$ and $\alpha$) the same as those represented by the blue dot (red square). The dimensionless parameters are obtained from the material constants presented in Table \ref{table:parameters}. In both cases (blue solid and red dotted lines), we see the general trend that the viscous effects become significant when the dimensionless parameter $\tilde{R}$ decreases. As the densities and sound speeds do not change significantly for different fluids, the value of $\tilde{R}$ is mainly determined by the radius $R$ of the vibrating particle and the shear viscosity $\eta$. As expected, the viscous effects therefore become significant typically when the vibrating structure decreases in size or when the shear viscosity of the fluid increases.

\begin{figure}[h]
\centering
 \includegraphics[width=1\textwidth]{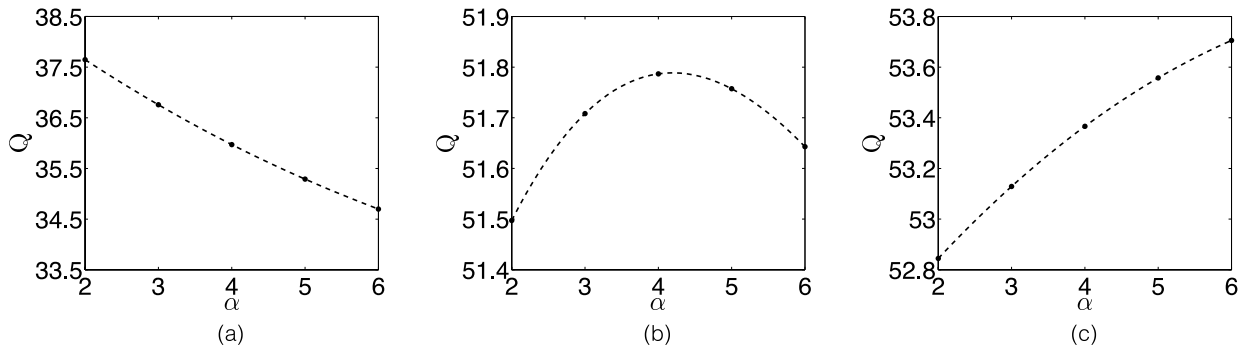}
        \caption{Parametric study of the quality factor $Q$ as a function of $\alpha$ for varying values of $\tilde{R}$. Dimensionless parameters $\tilde{\rho}$, $\tilde{c_t}$ and $\tilde{c}$ are obtained from material constants in Table \ref{table:parameters}. (a) $\tilde{R}=10$. (b) $\tilde{R}=50$. (c) $\tilde{R}=100$. }
        \label{fig:param_study}
\end{figure}

We also investigate the effect of the viscosity ratio $\alpha = \kappa/\eta$ in a compressible flow on the quality factor. Intuitively, the quality factor may decrease with a larger viscosity of the fluid because of the increased viscous dissipation. Such is the case for an incompressible flow. However, for the compressible flow considered here interesting variations are observed. In Fig.~\ref{fig:param_study}, we vary the viscosity ratio $\alpha$ at different values of $\tilde{R}=10, 50, 100$ (Fig.~\ref{fig:param_study}), keeping all other parameters fixed. As $\alpha$ increases, the quality factor can monotonically decrease (Fig.~\ref{fig:param_study}a), increase (Fig.~\ref{fig:param_study}c) or vary non-monotonically (Fig.~\ref{fig:param_study}b), depending on the value of $\tilde{R}$. This indicates that fluid compressibility provides a mechanism through which non-monotonic behaviors as a function of fluid viscosities are possible.

\subsection{Breathing mode of a gold nanosphere in a Maxwell fluid} \label{sec:viscoelastic}

The quality factor of high-frequency longitudinal oscillations of a single bipyramidal gold nanoparticle along its major axis in a water-glycerol mixture was recently measured by Pelton \textit{et al.} \cite{pelton} A non-monotonic dependence of the quality factor on the glycerol mass fraction $\chi$ in the mixture was observed: the quality factor first decreased as the glycerol mass fraction increased, reaching a minimum before increasing again. The longitudinal oscillation of the bipyramidal nanoparticle was modeled by dividing the bipyramid into infinitesimal sections along its major axis and approximating each section by a circular cylinder, where the solution for a longitudinally oscillating circular cylinder could be applied. \cite{chakraborty,pelton} By considering an incompressible flow and increasing glycerol mass fraction (hence, increasing the shear viscosity), a Newtonian fluid model predicted a monotonically decreasing relationship between the quality factor and the glycerol mass fraction, failing to account for the observed experimental dependence.  A Maxwell fluid model on the other hand captured the non-monotonic behavior. It was therefore concluded that the non-monotonic variation manifested the intrinsic viscoelastic properties of simple liquids.

For the bipyramidal geometry considered in Pelton \textit{et al.},\cite{pelton} the laser-induced excitation mechanism mainly excited the longitudinal vibration mode of the bipyramid. For nanospheres, the spherically symmetric fundamental breathing mode described in Sec.~\ref{sec:formulation} is excited instead. \cite{nano} Motivated by the experiment in Pelton \textit{et al.},\cite{pelton} we investigate how the quality factor of the breathing mode of an elastic sphere varies with the glycerol mass fraction in water-glycerol mixtures. The spherically symmetric geometry allows exact and analytical eigenvalue equations in both Newtonian (\ref{eqn:eigen_newton}) and viscoelastic fluid media.

\begin{figure}[h]
\centering
 \includegraphics[width=1\textwidth]{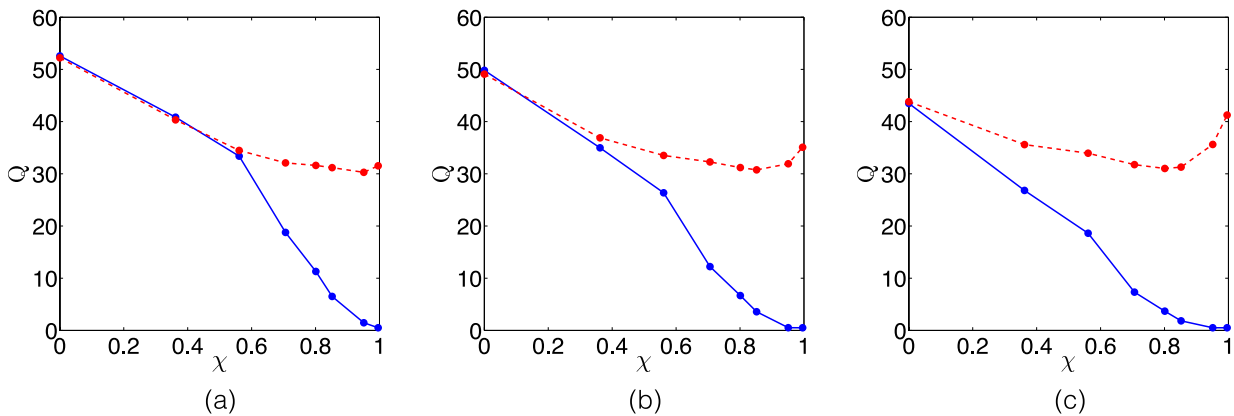}
        \caption{Quality factor $Q$ of radial vibrations of a sphere with radius $R$ as a function of glycerol mass-fraction $\chi$. (a) $R$ = 40 nm. (b) $R$ = 20 nm. (c) $R$ = 10 nm. The red dotted lines represent the Maxwell model; the blue solid lines represent the Newtonian model.}
        \label{fig:prl}
\end{figure}

With the same properties of water-glycerol mixtures at different mass fractions used in Pelton \textit{et al.}\cite{pelton, slie} (refer to Appendix \ref{sec:data} for a summary of these material properties),  we compute the eigenfrequency predicted by the Newtonian fluid model using (\ref{eqn:eigen_newton}). A monotonic decrease in the quality factor with the glycerol mass fraction $\chi$ is observed in Fig.~\ref{fig:prl} (blue lines) for nanospheres with radii of 10, 20, and 40 nm.

%\textcolor{blue}{Pelton \textit{et al.} \cite{pelton} obtained the material properties of the water-glycerol mixture for a wide range of glycerol mass fractions $\chi$ from the experimental work by Slie \textit{et al.}\cite{slie} (refer to Appendix \ref{sec:data} for a summary of the material constants). We use the same fluid properties and the Newtonian fluid model to compute the eigenfrequency  predicted by (\ref{eqn:eigen_newton}),} and observe a monotonic decrease in the quality factor with the glycerol mass fraction $\chi$ in Fig.~\ref{fig:prl} (blue lines) for nanospheres with radii of 10, 20, and 40 nm.

We employ the Maxwell model to describe the viscoelasticity of the fluid medium and take into account the effect of fluid compressibility in a similar fashion as Khismatullin and Nadim. \cite{ali} The total stress tensor $\sigmaB$ in a compressible viscoelastic fluid can be written as 
\begin{align}
\sigmaB = \left(-p + \kappa \nabla \cdot \v\right) \I+\tauB,
\end{align}
which represents the sum of the isotropic part and the deviatoric stress tensor $\tauB$ that has a vanishing trace for the linear viscoelastic case considered here.\cite{ali} In a Maxwell fluid, the viscoelastic behavior is modeled as a purely viscous damper and a purely elastic spring connected in series. Due to the series connection, the total deviatoric stress in the viscoelastic fluid $\tauB$ is the same as the deviatoric stress in the viscous damper $\tauB_\DT$ as well as that in the elastic spring $\tauB_\ST$, i.e.
\begin{align}
 \tauB = \tauB_\DT= \tauB_\ST . \label{eqn:stress}
\end{align}
However, the total rate of strain $\gammaB$ is a sum of the contributions from the damper $\gammaB_\DT$ and elastic spring $\gammaB_\ST$
\begin{align}
\gammaB = \gammaB_\DT+ \gammaB_\ST. \label{eqn:strain}
\end{align}
The shear stress in the viscous damper is given by a Newtonian constitutive equation 
\begin{align}
\tauB_\DT = 2 \eta_\DT \left( \dot{\gammaB}_{\DT} - \frac{\tr(\dot{\gammaB}_\DT)}{3} \I\right), \label{eqn:FluidShearStress}
\end{align}
where $\eta_\DT$ is the shear viscosity of the damper. The shear stress in the elastic spring is given by
\begin{align}
\tauB_\ST = 2 E \gammaB_\ST, \label{eqn:SolidShearStress}
\end{align}
where $E$ is the elastic modulus. We differentiate (\ref{eqn:strain}) with respect to time and use (\ref{eqn:FluidShearStress}) and (\ref{eqn:SolidShearStress}) to obtain a constitutive equation for the total shear stress $\tauB$ and total rate of strain $\dot{\gammaB}$
\begin{align}
\tauB + \lambda \dot{\tauB} = 2 \eta\left( \dot{\gammaB}- \frac{\tr{(\dot{\gammaB})}}{3}\I \right), \label{eqn:Viscoelastic}
\end{align}
where $\lambda = \eta/E$ is the relaxation time. In the derivation we have used that the total deviatoric shear stress is traceless, $\tr(\tauB)=0$, which implies $\tr(\tauB_{\ST})=0$ according to (\ref{eqn:stress}) and therefore $\tr(\gammaB_\ST)=0$  by (\ref{eqn:SolidShearStress}). As a result, we have $\tr(\dot{\gammaB}_\DT) = \tr(\dot{\gammaB}) $ in (\ref{eqn:Viscoelastic}). When the relaxation time is zero, $\lambda = 0$, (\ref{eqn:Viscoelastic}) reduces to the Newtonian constitutive equation (\ref{eqn:Newtonian}).
Since harmonic solutions are sought for the velocity field, the total shear stress tensor $\tauB$ should also have the same temporal dependence, $\exp(-i\omega t)$. We can therefore rewrite (\ref{eqn:Viscoelastic}), with this time dependence assumed, as
\begin{align}
\tauB = \frac{2\eta}{1-i\lambda \omega} \left(  \dot{\gammaB}- \frac{\tr(\dot{\gammaB})}{3} \I \right). \label{eqn:PViscoelsatic}
\end{align}

Comparing (\ref{eqn:PViscoelsatic}) with the Newtonian constitutive relation (\ref{eqn:Newtonian}), we observe that the eigenvalue equation for the breathing mode of an elastic sphere in a Maxwell fluid can be obtained simply by making the following transformation in (\ref{eqn:eigen_newton}), \cite{pelton} 
\begin{align}
\eta \rightarrow \frac{\eta}{1-i\lambda \omega} \cdot \label{eqn:transform}
\end{align}

Now using the relaxation times of water-glycerol mixtures at different glycerol mass fractions, \cite{slie} with the other parameters the same as in the Newtonian case in Fig.~\ref{fig:prl} (blue solid lines), we obtain the quality factor as a function of the glycerol mass fraction $\chi$ (Fig.~\ref{fig:prl}, red dotted lines) for different values of the nanosphere radius (Fig.~\ref{fig:prl}a, b, c: 40 nm, 20 nm, 10 nm, respectively). The results for the Newtonian and viscoelastic models agree when the glycerol mass fraction $\chi$ is smaller than a certain critical value, which depends on the radius of the nanosphere. More significantly, the viscoelastic model predicts a non-monotonic variation as a function of $\chi$, which is similar to the case of longitudinal vibration of bipyramidal gold nanoparticle reported in Pelton \textit{et al.}, \cite{pelton} even though the vibration mechanism is fundamentally different.

In Fig.~\ref{fig:prl}a ($R = 40$ nm), the Newtonian model predicts similar results as the viscoelastic model when $\chi \lesssim 0.56$ because the relaxation time of the mixture for these glycerol mass fractions is not large enough for the elastic effect to be significant. From (\ref{eqn:transform}) we can see that the elastic effects become significant if ${\vert \lambda \omega \vert \gtrsim 1}$, i.e. when the relaxation time scale of the fluid $\lambda$ is comparable with the vibration time scale $1/\omega_r$. Given a vibration frequency, we can therefore estimate the glycerol mass fraction beyond which the results of the viscoelastic model would deviate significantly from the Newtonian model. For instance, the vibration frequency ($f = \omega_r/2\pi$) of a gold nanosphere with $R=40$ nm in vacuum is given by (\ref{eqn:vacuum}) as $f \approx 38$ GHz, and it can be verified that the real part of the vibration frequency depends very weakly on the surrounding medium for low-damping vibrations considered here (see the values for different media in Table \ref{table:parameters}). The critical relaxation time such that $\vert\lambda_c \omega \vert = 1$ is given by $\lambda_c \approx 4$ ps, which occurs when $\chi \approx 0.56$ (refer to Table \ref{table:data_maxwell} for the relaxation times of different water-glycerol mixtures). When the radius of the gold nanosphere decreases to $R=20$ nm, the vibration frequency increases to $f\approx76$ GHz by (\ref{eqn:vacuum}), and the corresponding critical relaxation time decreases to $\lambda_c \approx 2$ ps. Hence the results of the viscoelastic model begin to deviate from that of the Newtonian model when $\chi \approx  0.36$ (Fig.~\ref{fig:prl}b). Similarly, the vibration frequency increases ($f \approx 152$ GHz) when the nanosphere radius is reduced to 10 nm, which gives a smaller critical relaxation time ($\lambda_c \approx 1$ ps), and hence a deviation from the Newtonian results at a smaller $\chi$ (Fig.~\ref{fig:prl}c).

\section{Discussion and Conclusion} \label{sec:discussion}

In this paper, we have revisited a classical calculation of the natural frequencies of a radially oscillating elastic sphere in simple and complex fluid media. We first considered the Newtonian fluid model taking into account both shear and bulk viscosities, and demonstrated that the fluid compressibility plays a significant role in the breathing mode of vibrating nanospheres. Should the limit of an incompressible flow be considered (see Appendix \ref{appendix:incompress}), the quality factor is significantly overestimated ($Q=841$), compared with the case of a compressible flow ($Q=52.6$). Although Pelton \textit{et al.} \cite{pelton} showed numerically that the effect of compressibility is insignificant for the case of longitudinal vibration of bipyramidal gold nanoparticles, we provide here an example where it is important to consider fluid compressibility in order to reasonably estimate the quality factor of the breathing mode of a nanosphere. Physically, the longitudinal vibration of the bipyramidal nanoparticle propagates through the shearing motion in the fluid, and hence the effect of fluid compressibility is insignificant. In contrast, due to the spherical symmetry in the breathing mode of a nanosphere, fluid compressibility plays a more significant role in propagating the energy through acoustic waves.

A viscoelastic response in a fluid triggered by high-frequency vibrations was demonstrated by Pelton \textit{et al.} \cite{pelton} for a longitudinal vibration of bipyramidal gold nanoparticles, where they observed non-monotonic variations of the quality factor as a function of glycerol mass fraction that were not captured by a Newtonian fluid model. Motivated by these experimental observations, we have extended the classical problem of a radially vibrating elastic sphere in a Newtonian fluid to a viscoelastic fluid, modeled as a Maxwell fluid. Despite the fundamental difference in the vibration mode, the breathing mode of a nanosphere, taking into account the relaxation time, also predicts a non-monotonic variation of the quality factor as a function of the glycerol mass fraction, similar to the response of longitudinal vibration of a gold bipyramidal nanoparticle in a Maxwell fluid. A Newtonian fluid model fails to capture this non-monotonic behavior. Due to the simplicity of the spherical geometry considered in this work, the eigenvalue equation for the breathing mode in the viscoelastic fluid medium is exact and analytical.

The relaxation time $\lambda$ of water-glycerol mixtures is typically small, on the order of 1--100 ps (Table \ref{table:data_maxwell}); the viscoelastic response is triggered only when the vibration frequency is sufficiently high such that ${\vert \lambda \omega \vert \gtrsim 1}$. For the cases studied in this paper, the real part of the vibration frequency can be effectively approximated by the vibration frequency of a sphere in vacuum (\ref{eqn:vacuum}), where we solve for $s=k_s R$ for a given ratio of wave speeds in the elastic solid $c_l/c_t$. One can rewrite the condition ${\vert \lambda \omega \vert \gtrsim 1}$ as ${\vert \lambda c_l s/ R \vert \gtrsim 1}$, which leads to the condition $R \lesssim \lambda c_l |s|$, a typical radius of the elastic structure smaller than which the viscoelastic response in the fluid would be triggered. For gold, $\vert s \vert \approx 3$ by (\ref{eqn:vacuum}), and hence the viscoelastic effect has to be taken into account when the radius of the gold nanosphere is smaller than $R\approx 35$ nm for a water-glycerol mixture with $\chi=0.56$, according to the material properties in Tables \ref{table:parameters} and \ref{table:data_maxwell}.

We also comment on the idea of destroying virus particles by acoustic resonance (see Introduction). The lifetime (or damping time, $\tau_D = -1/\omega_i$)  of the vibration was used in the literature \cite{ford,sav_murray, talati, stephanidis} to assess the feasibility of the idea. A major obstacle is the short damping time when the virus particle is embedded in a liquid. \cite{sav_murray, stephanidis} Assuming that the density and elastic properties of viruses were close to that of protein crystals (lysozyme), \cite{tachibana} the inviscid model by Talati and Jha \cite{talati} estimated that the damping time for a virus in liquid was of the order of picoseconds. \footnote{The density of lysozyme is $\rho_s=1210$ kg/$\text{m}^3$. The longitudinal and transverse wave speeds are $c_\ell =1817$ m/s and $c_t=915$ m/s, respectively. See Refs.~\cite{talati, tachibana} for details} Specifically, for a virus particle of 50 nm radius, Talati and Jha estimated a damping time of 17.3 ps for a virus-water configuration. A longer damping time (34.4 ps) was estimated for a virus-glycerol configuration, \cite{talati} leading to a conclusion that this configuration was more favorable than the virus-water configuration for virus destruction. Here we comment on the effect of fluid viscoelasticity on these estimations. For the virus-water configuration, the relaxation time of water is small compared with the vibration time scale ($\vert \lambda \omega \vert \approx 0.06$ in this case), and hence the effect of viscoelasticity is negligible. However, the viscoelastic effect is significant for the virus-glycerol configuration because the relaxation time of glycerol is comparable to the vibration time scale ($\vert \lambda \omega \vert \approx 37$). Our calculation shows that the viscoelastic response triggered in glycerol significantly increases the damping time to 111 ps, suggesting a better likelihood of destroying the virus compared with previous estimations.

%\textcolor{blue}{In previous works on the vibrational characteristics of viruses, the fluid medium was considered to be Newtonian.\cite{ford,sav_murray, talati} For example, Talati and Jha\cite{talati} modeled the virus as an elastic sphere ($\rho=1000$ kg/m$^3$, $c_\ell =1817$ m/s, $c_t=915$ m/s, $R=50$ nm) and considered its vibrations in water and glycerol media which were modeled as inviscid Newtonian fluids. They found that the damping time in glycerol ($34.4$ ps) was larger than that in water ($17.3$ ps), and therefore concluded that glycerol was a more favorable medium for virus destruction through resonant ultrasound waves. We use the same material parameters and study the effect of viscoelasticity on the damping of the virus. While the damping time in water changes only slightly due to viscoelasticity ($\mid \lambda_w \omega_w \mid \approx 0.06$), the damping time in glycerol increases up to $111$ ps ($\mid \lambda_g \omega_g \mid \approx 37 \gg 1$). This shows that the energy storing mechanism triggered in glycerol can significantly increase the lifetime of the vibrations of nanometer scale viruses, which is essential in the practice of virus destruction.}

Finally, we note that experimental measurements of the quality factor of nanoparticles are considerably lower than the theoretical predictions. Previous research attributed the discrepancy to other dissipation mechanisms intrinsic to the particle that were not taken into account in the theoretical model, for example, internal friction within the nanoparticle and the dissipation in the capping layer surrounding the nanoparticle. \cite{nano} Theoretical models taking into account these damping mechanisms and other plausible effects, such as the thermo-acoustic effect, are currently under investigation.

\section{Acknowledgements}

HAS thanks the Thermal Engineering Department of Tsinghua University for hosting a visit, where he first began to think about this problem. OSP thanks the Croucher Foundation for support through a Croucher Fellowship. VG is grateful to the Columbia Undergraduate Scholars Program for the Summer Enhancement Fellowship and to the Department of Mechanical and Aerospace Engineering at Princeton University for financial support and hospitality. We thank the NSF for partial support via grant CBET-1234500.

\appendix
\section{Newtonian Incompressible Fluid}
\label{appendix:incompress}

Consider small-amplitude waves in an incompressible viscous fluid described by the linearized Navier-Stokes equation
\begin{align}
\nabla \cdot \v &= 0, \label{eqn:InComCon}\\
\rho_f \frac{\partial \v}{\partial t} &= -\nabla p + \eta \nabla^2 \v . \label{eqn:navier_stokes_incompress}
\end{align}

Note that due to the spherical symmetry of the problem the identity $\nabla^2 \v = \nabla(\nabla \cdot \v) = \mathbf{0}$ holds, and  (\ref{eqn:navier_stokes_incompress}) simplifies to
\begin{align}
\rho_f \frac{\partial \v}{\partial t} &= -\nabla p . \label{eqn:navier_stokes_incompress_simple}
\end{align}

Assuming time-periodic oscillations, the velocity field $\v = v(r,t) \eh_r$ can be obtained from the continuity equation (\ref{eqn:InComCon}) as
\begin{align}
v(r,t) = \sum_{n=1}^\infty \frac{\tilde{B}_n}{r^2} e^{-i\omega_n t}, \label{eqn:incompress_vel_field}
\end{align}
where $\tilde{B}_n$ are arbitrary constants, and $\omega_n$ are the frequencies of normal modes. The pressure field in the fluid can then be determined from (\ref{eqn:navier_stokes_incompress_simple}) as
\begin{align}
p(r,t) &=  -\sum_{n=1}^\infty \tilde{B}_n \frac{i \rho_f \omega_n}{r} e^{-i \omega_n t} .
\end{align}

While the stress in the solid (\ref{eqn:SolidStress}) remains unchanged, the stress in the fluid becomes 
\begin{align}
\sigma_{rr}^f = -p + 2\eta \frac{\partial v}{\partial r} = \sum_{n=1}^\infty \tilde{B}_n \left( \frac{i \rho_f \omega_n}{r} -\frac{4\eta}{r^3} \right) e^{-i \omega_n t} . \label{eqn:fluid_stress_incompress}
\end{align}

We again couple the solid and fluid problems by matching the velocity and stress at the boundary, $r=R$, and requiring the existence of a non-trivial solution, which leads to (\ref{eqn:eigen_newton_incompress}) in the main text.

\section{Material parameters of the water-glycerol mixture}\label{sec:data}

The density, $\rho_f$, shear viscosity, $\eta$, bulk viscosity, $\kappa$, and speed of sound, $c$, of the water-glycerol mixture for different mole fractions of glycerol, $\chi_\nu$, were experimentally measured by Slie \textit{et al.}\cite{slie} We obtain the parameter values for different mass fractions of glycerol, $\chi$, by using the relation $\chi = \chi_\nu/(\chi_\nu + (1-\chi_\nu) \mu_w/\mu_g)$, where $\mu_w = 18$ g/mol and $\mu_g=92$ g/mol are the molar masses of pure water and pure glycerol respectively. We use the relation $\lambda = \eta/G_\infty$ and the measured values of the high frequency shear modulus, $G_\infty$, to find the relaxation time, $\lambda$, of the water-glycerol mixture for different values of $\chi$. \cite{slie, pelton} Numerical values of the parameters are summarized in Table \ref{table:data_maxwell}.

\begin{table} [h]
\scriptsize
\begin{tabular*}{\textwidth} { >{\centering\arraybackslash} m{1.5cm}  | >{\centering\arraybackslash} m{2.8cm} | >{\centering\arraybackslash} m{2.8cm} | >{\centering\arraybackslash} m{2.8cm} | >{\centering\arraybackslash} m{2.8cm} | >{\centering\arraybackslash} m{2.8cm} }
\hline

$\chi$ & $\rho_f$ (kg/m$^3$) & $\eta$ (Pa $\cdot$ s) & $\kappa$  (Pa $\cdot$ s) & $c$ (m/s) & $\lambda$ (ps) \\ \hline \hline

0	&	1000	&	0.000894	&	0.00286	&	1510	&	0.647	\\
0.36	&	1090	&	0.00270	&	0.00756	&	1662	&	1.87	\\
0.56	&	1140	&	0.00527	&	0.0116	&	1760	&	3.51	\\
0.71	&	1190	&	0.0200	&	0.0400	&	1830	&	12.7	\\
0.8	&	1210	&	0.0447	&	0.0760	&	1885	&	27.1	\\
0.85	&	1220	&	0.0923	&	0.120	&	1909	&	54.2	\\
0.95	&	1250	&	0.452	&	0.407	&	1920	&	243	\\
1	&	1260	&	0.988	&	0.790	&	1930	&	500	\\ \hline

\end{tabular*}
\caption{Values of fluid density ($\rho_f$), shear viscosity ($\eta$), bulk viscosity ($\kappa$), speed of sound ($c$) and relaxation time ($\lambda$) for different glycerol mass fractions ($\chi$).}
\label{table:data_maxwell}
\end{table}


\begin{thebibliography}{23}%
\makeatletter
\providecommand \@ifxundefined [1]{%
 \@ifx{#1\undefined}
}%
\providecommand \@ifnum [1]{%
 \ifnum #1\expandafter \@firstoftwo
 \else \expandafter \@secondoftwo
 \fi
}%
\providecommand \@ifx [1]{%
 \ifx #1\expandafter \@firstoftwo
 \else \expandafter \@secondoftwo
 \fi
}%
\providecommand \natexlab [1]{#1}%
\providecommand \enquote  [1]{``#1''}%
\providecommand \bibnamefont  [1]{#1}%
\providecommand \bibfnamefont [1]{#1}%
\providecommand \citenamefont [1]{#1}%
\providecommand \href@noop [0]{\@secondoftwo}%
\providecommand \href [0]{\begingroup \@sanitize@url \@href}%
\providecommand \@href[1]{\@@startlink{#1}\@@href}%
\providecommand \@@href[1]{\endgroup#1\@@endlink}%
\providecommand \@sanitize@url [0]{\catcode `\\12\catcode `\$12\catcode
  `\&12\catcode `\#12\catcode `\^12\catcode `\_12\catcode `\%12\relax}%
\providecommand \@@startlink[1]{}%
\providecommand \@@endlink[0]{}%
\providecommand \url  [0]{\begingroup\@sanitize@url \@url }%
\providecommand \@url [1]{\endgroup\@href {#1}{\urlprefix }}%
\providecommand \urlprefix  [0]{URL }%
\providecommand \Eprint [0]{\href }%
\providecommand \doibase [0]{http://dx.doi.org/}%
\providecommand \selectlanguage [0]{\@gobble}%
\providecommand \bibinfo  [0]{\@secondoftwo}%
\providecommand \bibfield  [0]{\@secondoftwo}%
\providecommand \translation [1]{[#1]}%
\providecommand \BibitemOpen [0]{}%
\providecommand \bibitemStop [0]{}%
\providecommand \bibitemNoStop [0]{.\EOS\space}%
\providecommand \EOS [0]{\spacefactor3000\relax}%
\providecommand \BibitemShut  [1]{\csname bibitem#1\endcsname}%
\let\auto@bib@innerbib\@empty
%</preamble>
\bibitem [{\citenamefont {Portales}\ \emph {et~al.}(2008)\citenamefont
  {Portales}, \citenamefont {Goubet}, \citenamefont {Saviot}, \citenamefont
  {Adichtchev}, \citenamefont {Murray}, \citenamefont {Mermet}, \citenamefont
  {Duval},\ and\ \citenamefont {Pileni}}]{portales}%
  \BibitemOpen
  \bibfield  {author} {\bibinfo {author} {\bibfnamefont {H.}~\bibnamefont
  {Portales}}, \bibinfo {author} {\bibfnamefont {N.}~\bibnamefont {Goubet}},
  \bibinfo {author} {\bibfnamefont {L.}~\bibnamefont {Saviot}}, \bibinfo
  {author} {\bibfnamefont {S.}~\bibnamefont {Adichtchev}}, \bibinfo {author}
  {\bibfnamefont {D.~B.}\ \bibnamefont {Murray}}, \bibinfo {author}
  {\bibfnamefont {A.}~\bibnamefont {Mermet}}, \bibinfo {author} {\bibfnamefont
  {E.}~\bibnamefont {Duval}}, \ and\ \bibinfo {author} {\bibfnamefont {M.-P.}\
  \bibnamefont {Pileni}},\ }\bibfield  {title} {\enquote {\bibinfo {title}
  {Probing atomic ordering and multiple twinning in metal nanocrystals through
  their vibrations},}\ }\href@noop {} {\bibfield  {journal} {\bibinfo
  {journal} {Proc. Natl. Acad. Sci. U.S.A.}\ }\textbf {\bibinfo {volume}
  {105}},\ \bibinfo {pages} {14784--14789} (\bibinfo {year}
  {2008})}\BibitemShut {NoStop}%
\bibitem [{\citenamefont {Jensen}, \citenamefont {Kim},\ and\ \citenamefont
  {Zettl}(2008)}]{jensen}%
  \BibitemOpen
  \bibfield  {author} {\bibinfo {author} {\bibfnamefont {K.}~\bibnamefont
  {Jensen}}, \bibinfo {author} {\bibfnamefont {K.}~\bibnamefont {Kim}}, \ and\
  \bibinfo {author} {\bibfnamefont {A.}~\bibnamefont {Zettl}},\ }\bibfield
  {title} {\enquote {\bibinfo {title} {An atomic-resolution nanomechanical mass
  sensor},}\ }\href@noop {} {\bibfield  {journal} {\bibinfo  {journal} {Nat.
  Nanotechnol.}\ }\textbf {\bibinfo {volume} {3}},\ \bibinfo {pages} {533--537}
  (\bibinfo {year} {2008})}\BibitemShut {NoStop}%
\bibitem [{\citenamefont {Verbridge}\ \emph {et~al.}(2006)\citenamefont
  {Verbridge}, \citenamefont {Bellan}, \citenamefont {Parpia},\ and\
  \citenamefont {Craighead}}]{verbridge}%
  \BibitemOpen
  \bibfield  {author} {\bibinfo {author} {\bibfnamefont {S.~S.}\ \bibnamefont
  {Verbridge}}, \bibinfo {author} {\bibfnamefont {L.~M.}\ \bibnamefont
  {Bellan}}, \bibinfo {author} {\bibfnamefont {J.~M.}\ \bibnamefont {Parpia}},
  \ and\ \bibinfo {author} {\bibfnamefont {H.~G.}\ \bibnamefont {Craighead}},\
  }\bibfield  {title} {\enquote {\bibinfo {title} {Optically driven resonance
  of nanoscale flexural oscillators in liquid},}\ }\href@noop {} {\bibfield
  {journal} {\bibinfo  {journal} {Nano Lett.}\ }\textbf {\bibinfo {volume}
  {6}},\ \bibinfo {pages} {2109--2114} (\bibinfo {year} {2006})}\BibitemShut
  {NoStop}%
\bibitem [{\citenamefont {Arlett}, \citenamefont {Myers},\ and\ \citenamefont
  {Roukes}(2011)}]{arlett}%
  \BibitemOpen
  \bibfield  {author} {\bibinfo {author} {\bibfnamefont {J.~L.}\ \bibnamefont
  {Arlett}}, \bibinfo {author} {\bibfnamefont {E.~B.}\ \bibnamefont {Myers}}, \
  and\ \bibinfo {author} {\bibfnamefont {M.~L.}\ \bibnamefont {Roukes}},\
  }\bibfield  {title} {\enquote {\bibinfo {title} {Comparative advantages of
  mechanical biosensors},}\ }\href@noop {} {\bibfield  {journal} {\bibinfo
  {journal} {Nat. Nanotechnol.}\ }\textbf {\bibinfo {volume} {6}},\ \bibinfo
  {pages} {203--215} (\bibinfo {year} {2011})}\BibitemShut {NoStop}%
\bibitem [{\citenamefont {Babincov\'{a}}, \citenamefont {Sourivong},\ and\
  \citenamefont {Babinec}(2000)}]{babincova}%
  \BibitemOpen
  \bibfield  {author} {\bibinfo {author} {\bibfnamefont {M.}~\bibnamefont
  {Babincov\'{a}}}, \bibinfo {author} {\bibfnamefont {P.}~\bibnamefont
  {Sourivong}}, \ and\ \bibinfo {author} {\bibfnamefont {P.}~\bibnamefont
  {Babinec}},\ }\bibfield  {title} {\enquote {\bibinfo {title} {Resonant
  absorption of ultrasound energy as a method of {HIV} destruction},}\ }\href
  {\doibase http://dx.doi.org/10.1054/mehy.2000.1088} {\bibfield  {journal}
  {\bibinfo  {journal} {Med. Hypotheses}\ }\textbf {\bibinfo {volume} {55}},\
  \bibinfo {pages} {450--451} (\bibinfo {year} {2000})}\BibitemShut {NoStop}%
\bibitem [{\citenamefont {Ford}(2003)}]{ford}%
  \BibitemOpen
  \bibfield  {author} {\bibinfo {author} {\bibfnamefont {L.~H.}\ \bibnamefont
  {Ford}},\ }\bibfield  {title} {\enquote {\bibinfo {title} {Estimate of the
  vibrational frequencies of spherical virus particles},}\ }\href@noop {}
  {\bibfield  {journal} {\bibinfo  {journal} {Phys. Rev. E}\ }\textbf {\bibinfo
  {volume} {67}},\ \bibinfo {pages} {051924} (\bibinfo {year}
  {2003})}\BibitemShut {NoStop}%
\bibitem [{\citenamefont {Saviot}\ and\ \citenamefont
  {Murray}(2004)}]{sav_murray}%
  \BibitemOpen
  \bibfield  {author} {\bibinfo {author} {\bibfnamefont {L.}~\bibnamefont
  {Saviot}}\ and\ \bibinfo {author} {\bibfnamefont {D.~B.}\ \bibnamefont
  {Murray}},\ }\bibfield  {title} {\enquote {\bibinfo {title} {Comment on
  "{E}stimate of the vibrational frequencies of spherical virus particles"},}\
  }\href@noop {} {\bibfield  {journal} {\bibinfo  {journal} {Phys. Rev. E}\
  }\textbf {\bibinfo {volume} {69}},\ \bibinfo {pages} {023901} (\bibinfo
  {year} {2004})}\BibitemShut {NoStop}%
\bibitem [{\citenamefont {Talati}\ and\ \citenamefont {Jha}(2006)}]{talati}%
  \BibitemOpen
  \bibfield  {author} {\bibinfo {author} {\bibfnamefont {M.}~\bibnamefont
  {Talati}}\ and\ \bibinfo {author} {\bibfnamefont {P.~K.}\ \bibnamefont
  {Jha}},\ }\bibfield  {title} {\enquote {\bibinfo {title} {Acoustic phonon
  quantization and low-frequency {R}aman spectra of spherical viruses},}\
  }\href {\doibase 10.1103/PhysRevE.73.011901} {\bibfield  {journal} {\bibinfo
  {journal} {Phys. Rev. E}\ }\textbf {\bibinfo {volume} {73}},\ \bibinfo
  {pages} {011901} (\bibinfo {year} {2006})}\BibitemShut {NoStop}%
\bibitem [{\citenamefont {Stephanidis}\ \emph {et~al.}(2007)\citenamefont
  {Stephanidis}, \citenamefont {Adichtchev}, \citenamefont {Gouet},
  \citenamefont {McPherson},\ and\ \citenamefont {Mermet}}]{stephanidis}%
  \BibitemOpen
  \bibfield  {author} {\bibinfo {author} {\bibfnamefont {B.}~\bibnamefont
  {Stephanidis}}, \bibinfo {author} {\bibfnamefont {S.}~\bibnamefont
  {Adichtchev}}, \bibinfo {author} {\bibfnamefont {P.}~\bibnamefont {Gouet}},
  \bibinfo {author} {\bibfnamefont {A.}~\bibnamefont {McPherson}}, \ and\
  \bibinfo {author} {\bibfnamefont {A.}~\bibnamefont {Mermet}},\ }\bibfield
  {title} {\enquote {\bibinfo {title} {Elastic properties of viruses},}\
  }\href@noop {} {\bibfield  {journal} {\bibinfo  {journal} {Biophys. J.}\
  }\textbf {\bibinfo {volume} {93}},\ \bibinfo {pages} {1354--1359} (\bibinfo
  {year} {2007})}\BibitemShut {NoStop}%
\bibitem [{\citenamefont {Hartland}(2006)}]{hartland}%
  \BibitemOpen
  \bibfield  {author} {\bibinfo {author} {\bibfnamefont {G.~V.}\ \bibnamefont
  {Hartland}},\ }\bibfield  {title} {\enquote {\bibinfo {title} {Coherent
  excitation of vibrational modes in metallic nanoparticles},}\ }\href@noop {}
  {\bibfield  {journal} {\bibinfo  {journal} {Annu. Rev. Phys. Chem.}\ }\textbf
  {\bibinfo {volume} {57}},\ \bibinfo {pages} {403--430} (\bibinfo {year}
  {2006})}\BibitemShut {NoStop}%
\bibitem [{\citenamefont {Ruijgrok}\ \emph {et~al.}(2012)\citenamefont
  {Ruijgrok}, \citenamefont {Zijlstra}, \citenamefont {Tchebotareva},\ and\
  \citenamefont {Orrit}}]{nano}%
  \BibitemOpen
  \bibfield  {author} {\bibinfo {author} {\bibfnamefont {P.~V.}\ \bibnamefont
  {Ruijgrok}}, \bibinfo {author} {\bibfnamefont {P.}~\bibnamefont {Zijlstra}},
  \bibinfo {author} {\bibfnamefont {A.~L.}\ \bibnamefont {Tchebotareva}}, \
  and\ \bibinfo {author} {\bibfnamefont {M.}~\bibnamefont {Orrit}},\ }\bibfield
   {title} {\enquote {\bibinfo {title} {Damping of acoustic vibrations of
  single gold nanoparticles optically trapped in water},}\ }\href@noop {}
  {\bibfield  {journal} {\bibinfo  {journal} {Nano Lett.}\ }\textbf {\bibinfo
  {volume} {12}},\ \bibinfo {pages} {1063--1069} (\bibinfo {year}
  {2012})}\BibitemShut {NoStop}%
\bibitem [{\citenamefont {Fujii}\ \emph {et~al.}(1991)\citenamefont {Fujii},
  \citenamefont {Nagareda}, \citenamefont {Hayashi},\ and\ \citenamefont
  {Yamamoto}}]{fujii}%
  \BibitemOpen
  \bibfield  {author} {\bibinfo {author} {\bibfnamefont {M.}~\bibnamefont
  {Fujii}}, \bibinfo {author} {\bibfnamefont {T.}~\bibnamefont {Nagareda}},
  \bibinfo {author} {\bibfnamefont {S.}~\bibnamefont {Hayashi}}, \ and\
  \bibinfo {author} {\bibfnamefont {K.}~\bibnamefont {Yamamoto}},\ }\bibfield
  {title} {\enquote {\bibinfo {title} {Low-frequency {R}aman scattering from
  small silver particles embedded in {S}i{O}2 thin films},}\ }\href@noop {}
  {\bibfield  {journal} {\bibinfo  {journal} {Phys. Rev. B}\ }\textbf {\bibinfo
  {volume} {44}},\ \bibinfo {pages} {6243--6248} (\bibinfo {year}
  {1991})}\BibitemShut {NoStop}%
\bibitem [{\citenamefont {Lamb}(1882)}]{lamb}%
  \BibitemOpen
  \bibfield  {author} {\bibinfo {author} {\bibfnamefont {H.}~\bibnamefont
  {Lamb}},\ }\bibfield  {title} {\enquote {\bibinfo {title} {On the vibrations
  of an elastic sphere},}\ }\href@noop {} {\bibfield  {journal} {\bibinfo
  {journal} {Proc. London Math. Soc.}\ }\textbf {\bibinfo {volume} {13}},\
  \bibinfo {pages} {189--212} (\bibinfo {year} {1882})}\BibitemShut {NoStop}%
\bibitem [{\citenamefont {Dubrovskiy}\ and\ \citenamefont
  {Morochnik}(1981)}]{dub}%
  \BibitemOpen
  \bibfield  {author} {\bibinfo {author} {\bibfnamefont {V.~A.}\ \bibnamefont
  {Dubrovskiy}}\ and\ \bibinfo {author} {\bibfnamefont {V.~S.}\ \bibnamefont
  {Morochnik}},\ }\bibfield  {title} {\enquote {\bibinfo {title} {Natural
  vibrations of a spherical inhomogeneity in an elastic medium},}\ }\href@noop
  {} {\bibfield  {journal} {\bibinfo  {journal} {Izv. Earth Phys.}\ }\textbf
  {\bibinfo {volume} {17}},\ \bibinfo {pages} {494--504} (\bibinfo {year}
  {1981})}\BibitemShut {NoStop}%
\bibitem [{\citenamefont {Kheisin}(1967)}]{kheisin}%
  \BibitemOpen
  \bibfield  {author} {\bibinfo {author} {\bibfnamefont {D.~E.}\ \bibnamefont
  {Kheisin}},\ }\bibfield  {title} {\enquote {\bibinfo {title} {Radial
  oscillations of an elastic sphere in a compressible fluid},}\ }\href@noop {}
  {\bibfield  {journal} {\bibinfo  {journal} {Fluid Dyn.}\ }\textbf {\bibinfo
  {volume} {2}},\ \bibinfo {pages} {53--55} (\bibinfo {year}
  {1967})}\BibitemShut {NoStop}%
\bibitem [{\citenamefont {Saviot}, \citenamefont {Netting},\ and\ \citenamefont
  {Murray}(2007)}]{sav}%
  \BibitemOpen
  \bibfield  {author} {\bibinfo {author} {\bibfnamefont {L.}~\bibnamefont
  {Saviot}}, \bibinfo {author} {\bibfnamefont {C.~H.}\ \bibnamefont {Netting}},
  \ and\ \bibinfo {author} {\bibfnamefont {D.~B.}\ \bibnamefont {Murray}},\
  }\bibfield  {title} {\enquote {\bibinfo {title} {Damping by bulk and shear
  viscosity of confined acoustic phonons for nanostructures in aqueous
  solution},}\ }\href@noop {} {\bibfield  {journal} {\bibinfo  {journal} {J.
  Phys. Chem. B}\ }\textbf {\bibinfo {volume} {111}},\ \bibinfo {pages}
  {7457--7461} (\bibinfo {year} {2007})}\BibitemShut {NoStop}%
\bibitem [{\citenamefont {Chakraborty}\ \emph {et~al.}(2013)\citenamefont
  {Chakraborty}, \citenamefont {van Leeuwen}, \citenamefont {Pelton},\ and\
  \citenamefont {Sader}}]{chakraborty}%
  \BibitemOpen
  \bibfield  {author} {\bibinfo {author} {\bibfnamefont {D.}~\bibnamefont
  {Chakraborty}}, \bibinfo {author} {\bibfnamefont {E.}~\bibnamefont {van
  Leeuwen}}, \bibinfo {author} {\bibfnamefont {M.}~\bibnamefont {Pelton}}, \
  and\ \bibinfo {author} {\bibfnamefont {J.~E.}\ \bibnamefont {Sader}},\
  }\bibfield  {title} {\enquote {\bibinfo {title} {Vibration of nanoparticles
  in viscous fluids},}\ }\href@noop {} {\bibfield  {journal} {\bibinfo
  {journal} {J. Phys. Chem. C}\ }\textbf {\bibinfo {volume} {117}},\ \bibinfo
  {pages} {8536--8544} (\bibinfo {year} {2013})}\BibitemShut {NoStop}%
\bibitem [{\citenamefont {Pelton}\ \emph {et~al.}(2013)\citenamefont {Pelton},
  \citenamefont {Chakraborty}, \citenamefont {Malachosky}, \citenamefont
  {Guyot-Sionnest},\ and\ \citenamefont {Sader}}]{pelton}%
  \BibitemOpen
  \bibfield  {author} {\bibinfo {author} {\bibfnamefont {M.}~\bibnamefont
  {Pelton}}, \bibinfo {author} {\bibfnamefont {D.}~\bibnamefont {Chakraborty}},
  \bibinfo {author} {\bibfnamefont {E.}~\bibnamefont {Malachosky}}, \bibinfo
  {author} {\bibfnamefont {P.}~\bibnamefont {Guyot-Sionnest}}, \ and\ \bibinfo
  {author} {\bibfnamefont {J.~E.}\ \bibnamefont {Sader}},\ }\bibfield  {title}
  {\enquote {\bibinfo {title} {Viscoelastic flows in simple liquids generated
  by vibrating nanostructures},}\ }\href@noop {} {\bibfield  {journal}
  {\bibinfo  {journal} {Phys. Rev. Lett.}\ }\textbf {\bibinfo {volume} {111}},\
  \bibinfo {pages} {244502} (\bibinfo {year} {2013})}\BibitemShut {NoStop}%
\bibitem [{Note1()}]{Note1}%
  \BibitemOpen
  \bibinfo {note} {The values of material parameters used in Ruijgrok \protect
  \textit {et al.}\cite {nano} are slightly different from those in Table \ref
  {table:parameters}, which result in the slight differences in the calculated
  quality factors}\BibitemShut {NoStop}%
\bibitem [{\citenamefont {Slie}, \citenamefont {Donfor},\ and\ \citenamefont
  {Litovitz}(1966)}]{slie}%
  \BibitemOpen
  \bibfield  {author} {\bibinfo {author} {\bibfnamefont {W.~M.}\ \bibnamefont
  {Slie}}, \bibinfo {author} {\bibfnamefont {A.~R.}\ \bibnamefont {Donfor}}, \
  and\ \bibinfo {author} {\bibfnamefont {T.~A.}\ \bibnamefont {Litovitz}},\
  }\bibfield  {title} {\enquote {\bibinfo {title} {Ultrasonic shear and
  longitudinal measurements in aqueous glycerol},}\ }\href@noop {} {\bibfield
  {journal} {\bibinfo  {journal} {J. Chem. Phys.}\ }\textbf {\bibinfo {volume}
  {44}},\ \bibinfo {pages} {3712--3718} (\bibinfo {year} {1966})}\BibitemShut
  {NoStop}%
\bibitem [{\citenamefont {Khismatullin}\ and\ \citenamefont
  {Nadim}(2002)}]{ali}%
  \BibitemOpen
  \bibfield  {author} {\bibinfo {author} {\bibfnamefont {D.}~\bibnamefont
  {Khismatullin}}\ and\ \bibinfo {author} {\bibfnamefont {A.}~\bibnamefont
  {Nadim}},\ }\bibfield  {title} {\enquote {\bibinfo {title} {Radial
  oscillations of encapsulated microbubbles in viscoelastic liquids},}\
  }\href@noop {} {\bibfield  {journal} {\bibinfo  {journal} {Phys. Fluids}\
  }\textbf {\bibinfo {volume} {14}},\ \bibinfo {pages} {3534--3557} (\bibinfo
  {year} {2002})}\BibitemShut {NoStop}%
\bibitem [{\citenamefont {Tachibana}\ \emph {et~al.}(2000)\citenamefont
  {Tachibana}, \citenamefont {Kojima}, \citenamefont {Ikuyama}, \citenamefont
  {Kobayashi},\ and\ \citenamefont {Ataka}}]{tachibana}%
  \BibitemOpen
  \bibfield  {author} {\bibinfo {author} {\bibfnamefont {M.}~\bibnamefont
  {Tachibana}}, \bibinfo {author} {\bibfnamefont {K.}~\bibnamefont {Kojima}},
  \bibinfo {author} {\bibfnamefont {R.}~\bibnamefont {Ikuyama}}, \bibinfo
  {author} {\bibfnamefont {Y.}~\bibnamefont {Kobayashi}}, \ and\ \bibinfo
  {author} {\bibfnamefont {M.}~\bibnamefont {Ataka}},\ }\bibfield  {title}
  {\enquote {\bibinfo {title} {Sound velocity and dynamic elastic constants of
  lysozyme single crystals},}\ }\href@noop {} {\bibfield  {journal} {\bibinfo
  {journal} {Chem. Phys. Lett.}\ }\textbf {\bibinfo {volume} {332}},\ \bibinfo
  {pages} {259--264} (\bibinfo {year} {2000})}\BibitemShut {NoStop}%
\bibitem [{Note2()}]{Note2}%
  \BibitemOpen
  \bibinfo {note} {The density of lysozyme is $\rho _s=1210$ kg/$\protect \text
  {m}^3$. The longitudinal and transverse wave speeds are $c_\ell =1817$ m/s
  and $c_t=915$ m/s, respectively. See Refs.~\cite {talati, tachibana} for
  details}\BibitemShut {NoStop}%
\end{thebibliography}
\end{document}